\documentclass[a4paper,11pt]{article}
\pdfoutput=1 

\usepackage{jinstpub} 
\usepackage{subfig}
\usepackage{wrapfig}
\usepackage[mathscr]{eucal}

\title{\boldmath RPC performance vs FE electoronics and detector parameters}

\author[a,1]{R.~Cardarelli,%
\note{Corresponding author.}}

\author[a]{and G.~Aielli,}
\author[a]{E.~Alunno Camelia,}
\author[a]{S.~Bruno,}
\author[a]{A.~Caltabiano,}
\author[a]{P.~Camarri,}
\author[a]{A.~Di Ciaccio,}
\author[b]{B.~Liberti,}
\author[b]{L.~Massa,}
\author[a]{L.~Pizzimento,}
\author[a]{A.~Rocchi.}

\affiliation[a]{INFN Rome Tor Vergata,\\
Via della Ricerca Scientifica 1, Rome, Italy}

\emailAdd{roberto.cardarelli@roma2.infn.it}

\abstract{}

\keywords{Resistive-plate chambers, Gaseous detectors,
Materials for gaseous detectors, Detector design and construction technologies and materials. 
}

\collaboration[c]{}

\proceeding{14$^{\text{th}}$ Workshop on Resistive Plate Chambers and Related Detectors\\
  19-23 February 2018\\
  Puerto Vallarta, Jalisco State, MEXICO}

\begin{document}
\maketitle
\flushbottom

\section{Introduction}
The first Resistive Plate Chambers detectors \cite{a}\cite{b} were developed for cosmic ray experiments, where low rate capability, good time resolution and low cost per unit of area were needed. These same features, except for the low rate capability, were required in the muon spectrometers of the collider experiments like LHC \cite{c}. For this purpose new RPC detectors with increased rate capability were developed. The rate capability improvement has been achieved thanks to the transition from streamer to saturated avalanche regime \cite{d}, in which the average charge produced in the gas dicharge is smaller. The price to pay for this working mode switch is the transfer of the amplification from the detector to the FE electorinics. 
The High luminosity LHC and the future colliders will require even greater rate capability in the muon spectrometer compared to the current one \cite{c}. For this reason, further improving of the rate capability  is required. 

The transition from the saturated avalanche to a low saturated avalanche regime moves in this direction and needs a new front end electronics with better signal to noise ratio, because the average induced charge on the electrode is even smaller. 
The front end electronics design is crucial for the RPC performances \cite{g}. In this paper we discuss the performances of the RPC detector changing the front end design and the detector parameters.

\section{Rate capability }
When a charged-particle flux $\Phi$ crosses an RPC detector, the overlapping of many processes cause a voltage drop on the electrodes. 
the voltage on the gas gap is 

\begin{equation}
 \label{equation}
 V_{gas}=V-2\rho d \langle Q \rangle \Phi_{eff}
\end{equation}

\noindent where $\rho$ and $d$ are respectivly the electrode resistivity e thickness, $\langle Q\rangle$ is the average charge involved in a single discharge process and $\Phi_{eff}$ is the average number of processes occurring in the detector per unit time and surface. 
Negletting the spontaneus random processes, that is related only to the detector build quality, the effective flux $\Phi_{eff}$ is a fraction $\epsilon\Phi$ of the particle flux, where $\epsilon$ is the efficiency that depends on the gas voltage $V_{gas}$.
 To prevent the detector from losing efficiency as $\Phi$ rises, it is necessary to minimize the average voltage drop on the electrodes in such a way to fix the $V_{gas}$ value. For this purpose the parameters $\rho$, $d$ an $Q$ that multiply the flux in equation \ref{equation} must be reduced.
 
To increase the rate capability it is chosen to move the amplification from the gas to the front end electronics leaving the electrod resistivity unchanged \cite{f}. The front end electronics is designed to maximize the signal to noise ratio and the detector performances taking into account the induced signal properties.

\section{Induced signal}

The charge produced during the discharge process in a RPC detector (and so the induced signal) changes drastically depending on the detector working mode. 
The Shockley-Ramo theorem allows to calculate the current $\mathtt{i_{k}}$ induced on an electrode by a moving particle $k$ with charge $e$ as  

\begin{equation}
 \label{Equation1}
 \mathtt{i}_{k} = e \vec{v_{k}}\cdot\vec{E_{w}}
\end{equation}

\noindent where $v_{k}$ is the carrier drift velocity and $E_{w}$ is the weighting field. 
In the discharge process both electrons and ions are generated. The total instantaneus induced current is given by summing  the equation \ref{Equation1} on all the envolved particles. The variations on the amount of charges during the process evolution can be descibed introducing as instantaneus drift velocity of each particles a step function defined as

\begin{equation}
 \label{Equation1.1}
 v_{k}(t) =v_{k}\bigl[\mathtt{H}\bigl(t-t_{k}^{start}\bigr)-\mathtt{H}\bigl(t-t_{k}^{stop}\bigr)\bigr]
\end{equation}\\

\noindent where $\mathtt{H}(t)$ is the Heaviside step function, $t_{k}^{start}$ is the paticle termalization time and $t_{k}^{stop}$ is the time at wich the particle reach the electrode ($t_{k}^{stop}>t_{k}^{start}>0$).
Semplifyng the pick-up electrodes for the signal readout as an infinite plane, the weighting field reduce to $1\; V/m$ and the equation \ref{Equation1} summation gives

\begin{equation}
 \label{Equation1.2}
 \mathtt{i}(t) =e \sum_{k=1}^{2N}{v_{k}}(t)= e \biggl(\sum_{i=1}^{N}{v_{i}}(t)+\sum_{e=1}^{N}{v_{e}}(t)\biggr)=\mathtt{i}_i(t)+\mathtt{i}_e(t)
\end{equation}\\

\noindent where $v_{i}(t)$ and ${v}_{e}(t)$ are the ionic and electronic instantaneus drift velocity respectively.
The discharge process nature has an intrinsic asymmetry: the discharge grows in the electrons drift direction, so most of the ions drift for a path $d_i$ longer than the electrons path $d_e$. The two carriers, moreover, differs in mass, and so in their drift velocity $v_k$, about five orders of magnitude. For this reasons the current pulse $\mathtt{i}_e(t)$ induced by electrons is five orders of magnitude higher than the current pulse $\mathtt{i}_i(t)$ induced by ions. The pulse duration time is directly proporsional to the carrier path $d_k$ and inversely proportional to the dtift velocity $v_k$. Experimental observations shows that the ionic pulse duration time is four order of magnitude longer then the electronic pulse so the electrons drift on avarege for a tenth of the ions path. This asimmetry has an effect on the total charge induced by the two different carriers. Infact suppose that each single carrier $k$ drifts across the gas gap for a distance $d_k$ with constant velocity $v_k$, the integration of \eqref{Equation1} reduce to

\begin{equation}
 \label{Equation2}
 \mathtt{Q_{k}} = e \frac{d_{k}}{D}
\end{equation}\\

\noindent where $\mathtt{Q_{k}}$ is the charge induced by the particle $k$ on the electrode and $D$ is the distance between the two pick-up electrodes. In order to highlight the induced signal properties, it is useful to separate the two carriers contributions. The total induced charge $\mathtt{Q}$ is simply given by summing the equation \ref{Equation2} on all the involved $2N$ charges. Introducing the average drift distances $\bar{d}_{e}$ and $\bar{d}_{i}$  it gives

\begin{equation}
 \label{Equation3}
 \mathtt{Q}=\mathtt{Q_e}+\mathtt{Q_i} = \frac{e}{D} \sum_{k=1}^{2N} d_{k}= \frac{eN}{D}(\bar{d}_{i}+\bar{d}_{e}) ; .\\
\end{equation}
 
The charge fraction induced by the electrons is smaller then the fraction induced by ions and it is as much small as the average electron drift distance $\bar{d}_{e}$ decrease.   

\begin{figure}[h]
\centering
{\includegraphics[trim=0.cm 0.cm 0cm 0.cm, clip=true,width=1\textwidth]{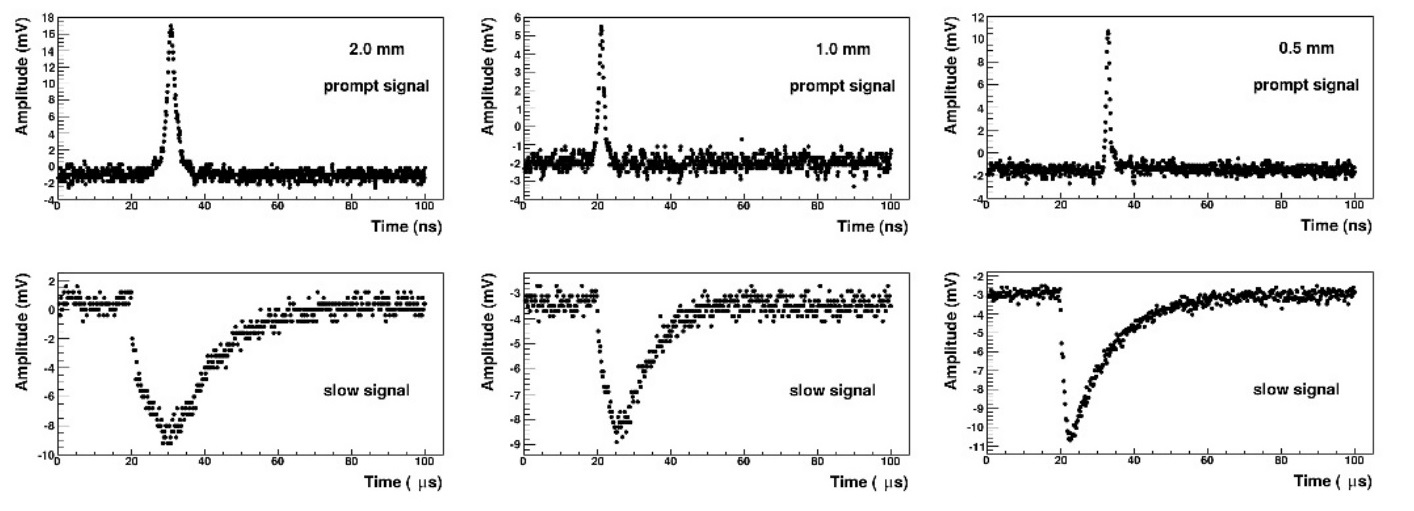}} \quad
\caption{\emph{Induced signals for three different gas gap thickness. The prompt signals, on the top and the ionic signals, on the bottom, are read on $50\;\Omega$ and $1\;M\Omega$ resectively. The prompt signal duration time decreases as the gas gap decreases \cite{f}. }}
\label{fig:figure1}
\end{figure}

In conclusion equations \ref{Equation1} and \ref{Equation3} show that the total induced signal can be separated into two components with very different property: the electronic component, \textit{prompt signal}, and the ionic component, \textit{ionic signal}. The first, shorter in duration, has higher amplitude and can be read on the external peak-up electrodes. The second, longer in duration, has smaller amplitude. In figure \ref{fig:figure1} the induced signals for three differents gas gap thickness are presented \cite{f}.

\section{Detector working mode and front end electronics parameters}
As described above, the induced signal depends on how many charges are produced during the discharge and how long and fast they drift inside the gas gap. This parameters are strictly related to the electric field inside the gas and to the gas mixture, therefore to the detector working mode. 

When The first RPC detector was developed, only the streamer working mode was known. In this regime the electric field on the gas gap is so high that the ion-electron pairs in the space charge avalanche region recombine emitting photons. These photons gives rise to an isotropic generation of avalanches, with the production of large amount of charges ($0.1\div 1\; nC$) along the whole thickness of the gas gap. The gas gain is defined as the ratio between the total charge and the charge produced by the primary ionization and take a value between $6\times10^7$ and $6\times10^8$ \cite{a}\cite{b}.  
 
The improvement of the front end electronics allowed the discovery of the avalanche discharge process \citep{d}. Study on the gas mixtures have shown that with the addiction of a high electronegative component, like $SF_6$, the avalanche process can be separated from the streamer process lowering the gas electric field \cite{e}.   
 In the avalanche working mode the discharge grows exponentially until the space charge effect sets in and the slow electrons are catched by the electronegative gas component. From this moment the total charge tends towards a costant value. In this situation the photons production is suppressed and the discharge evolves only in the electrons drift direction. This regime could be divided into two intervals depending on the electric field: the high saturated avalanche, in which the total charge take values between $(20\div40)\; pC$ and the low saturated avalanche, in which the total charge decreases until $(1\div5)\; pC$. Respect to the streamer working mode, the gas gain reduces by a factor greater then two order of magnitudes that must be recovered on the electronics front end.
 
\begin{figure}[h]
\centering
{\includegraphics[trim=0.cm 0.cm 0cm 0.cm, clip=true,width=1\textwidth]{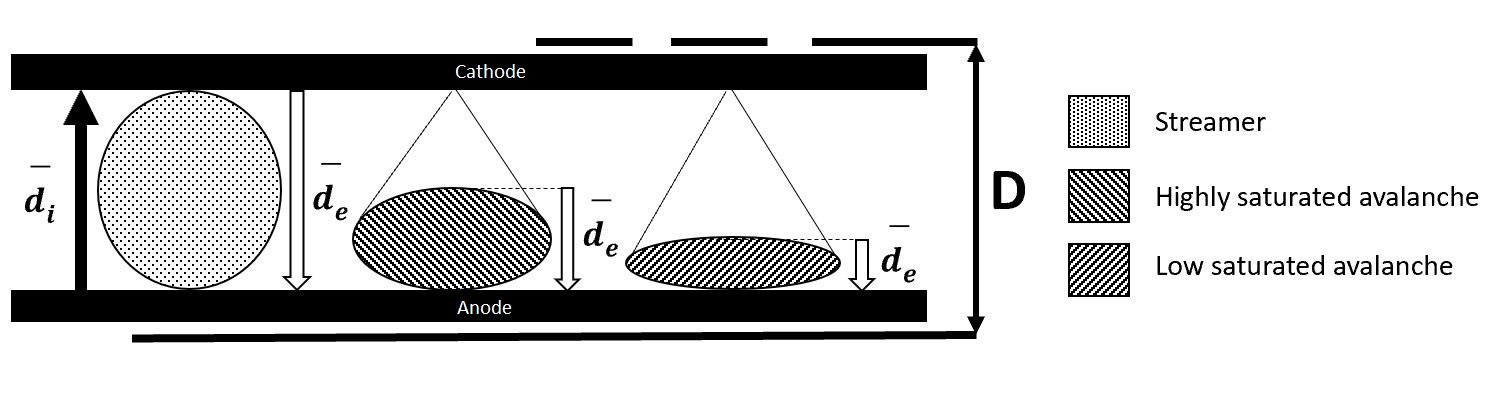}} \quad
\caption{\emph{Schematic representation of the discharge process evolution in the three known working modes. The ionic average drift distance does not change effectively, the electrons average drift distance, instead, reduces drastically. The electrodes thickness put an intrinsic limit on the collectible charge.}}
\label{fig:figure2}
\end{figure} 

The sketch in figure \ref{fig:figure2} represents the average charge distribution and the drift distance in different working modes. In streamer mode the discharge process grows in both the directions, so a significant amount of electrons drift for almost all the gap thickness and the concerning ionic to prompt charge ratio is of the order of $2\div5$. In high saturated avalanche mode the detector intrinsic asymmetry makes significant: the charge distribution is closer to the anode and the prompt to ionic charge ratio increase until $10\div20$. Lowering the electric field the charge distribution becomes sharper near the anode (low saturated avalanche regime). The ions average drift distance reaches the gas gap thickness and $\mathtt{Q_i}\approx Ne$. In this case the ionic to prompt charge ratio gives a very good estimation of the fraction of charge induced by the electrons respect to the electrons total charge and has values between $20\div30$ . 

\begin{figure}[h]
\centering
{\includegraphics[trim=0.cm 0.cm 0cm 0.cm, clip=true,width=0.7\textwidth]{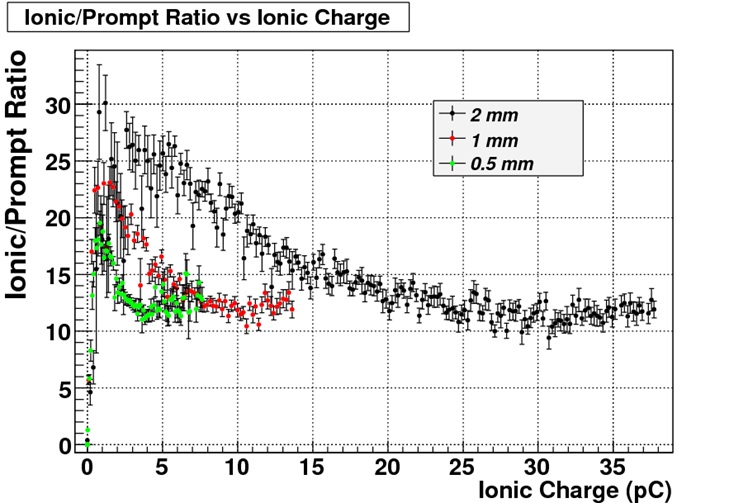}} \quad
\caption{\emph{Ionic to prompt induced charge ratio as a function of the ionic induced charge for three different gas gap thickness \cite{f}.}}
\label{fig:figure3}
\end{figure}

In figure \ref{fig:figure3} the ionic to prompt charge ratio is plotted as a function of the ionic induced charge for three different gas gap thickness \cite{f}. When the ionic charge is small, the ionic to prompt charge ratio has its maximum value. This is consistent with an interpretation in which the avalanche occurs near the anode, so the avalanche process stops quickly, the electrons drift for a short path and the ions taverse all the gas gap thickness.  When the ionic charge is high the ionic to prompt charge ratio is low because the discharge take place far from the anode and the large amount of carriers, produced in the avalanche moltiplication, drift for a longer path. The thinner the gas gap, the smaller the path the electrons could cross, and so the multiplication factor. Settled a threshold value on the induced signal discrimination, a reduction on the total gain involves a cut on all those processes that take place near the anode. This is the reason why the ionic to prompt charge ratio reaches first the saturation value in the thinner gas gap. 

To decide what is the upper limit front end electronics noise value, below which it's possible to work with a total charge $Q$, we need to take into account
  
\begin{itemize}
\item the ionic to prompt charge ratio;
\item the charge distribution;
\item the confidence interval of noise rejection. 
\end{itemize}

\begin{figure}[h]
\centering
{\includegraphics[trim=0.cm 0.cm 0cm 0.cm, clip=true,width=0.7\textwidth]{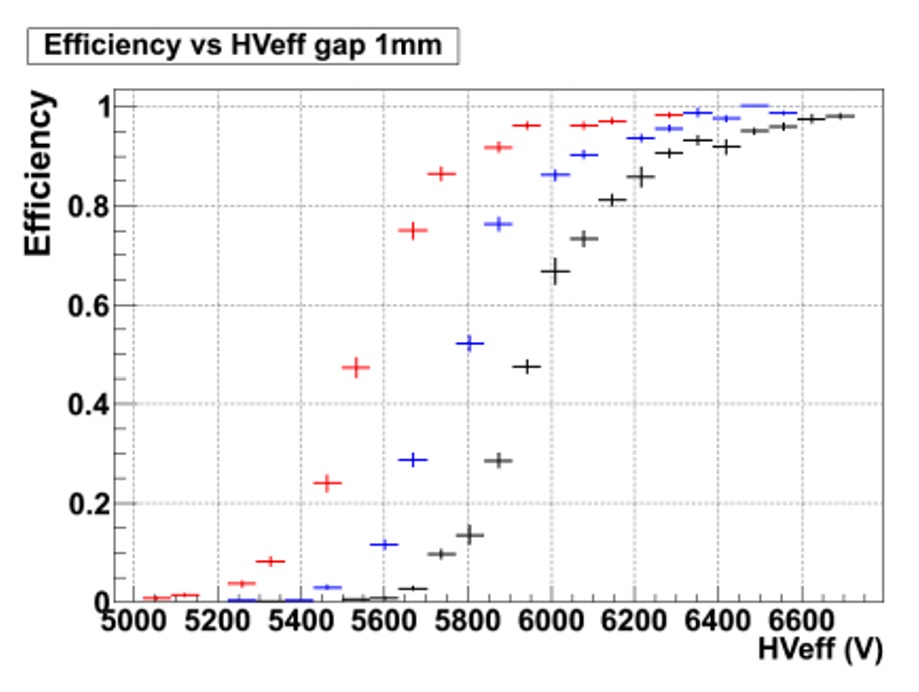}} \quad
\caption{\emph{Efficiency curves for a detector with $1 mm$ thick gas gap. In black is plotted the efficiency measured whitout preamplifier, in blu that with the preamplifer based on silicon technology and in red that measured with the preamplifier based on silicon-germanium technology.}}
\label{fig:figure4}
\end{figure}

\noindent In the saturated avalanche mode the ionic to prompt charge ratio is $25$, with a total electronics charge of the order of $(1\div5)\;pC$, this means that at most a charge of $(40\div80)\;fC$ is induced on the peak-up electrodes. 
The signal charge of an RPC detector has a distribution such that the discrimination threshold should be set at 1/10 of the average value, in way to have full efficiency. The noise counts are proportional to the product $BW\times[1-\Phi(x)]$, where $BW$ is the close-loop bendwidth and $\Phi(x)$ is the \textit{Cumulative Distribution Function} of the noise amplitude. To minimize the noise counts the threshold must be set at $5\times RMS_{noise}$. Considering these three factors, the front end electronics for an RPC detector must have a noise three order of magnitude lower then the total electronic charge. For an RPC detector working in low saturated avalanche mode this result in $10^4$ electrons of noise \cite{g}.
As regards the front end electronics bandwidth, it is related to the growth time of the avalanche discharge, that depends on the working mode. In the avalanche mode the prompt signal rise time $t_e$ decreases making the gas gap thiner, as figure \ref{fig:figure1} shows. The front end electronics requires an open loop bandwidth much greater than $\frac{3}{t_e}$. This requires an open loop bandwith greater than $600\;MHz$ for $2\;mm$ and $1GHz$ for $1\;mm$ gas gap thickness\cite{g}.
This performances represents a technological hurdle, becouse the preamplifer noise increases as the bandwith increases. That's why the detector physics knowledge evolves along toghter with the electronics improvements.
As an example of this, in figure \ref{fig:figure4} the efficiencies of a $1\;mm$ thick gas gap with different front end electronics are plotted as a function of the high voltage. The silicon-germanium heterojunction
technology allows to drop off the high voltage by about $600\;V$ without lose of efficiency. This means that the charge produced in the discharge process reduses by an order of magnitude and, further an improvement of the detector lifetime,  the rate capability increases up to $10\;kHz/cm^2$ \citep{c}.

\section{The time resolution question}
\begin{figure}[h]
\centering
\subfloat[][\emph{Time resolution as function of the gas gap thickness.}]
{\includegraphics[trim=0.cm 0.cm 0cm 0.cm, clip=true,width=0.45\textwidth]{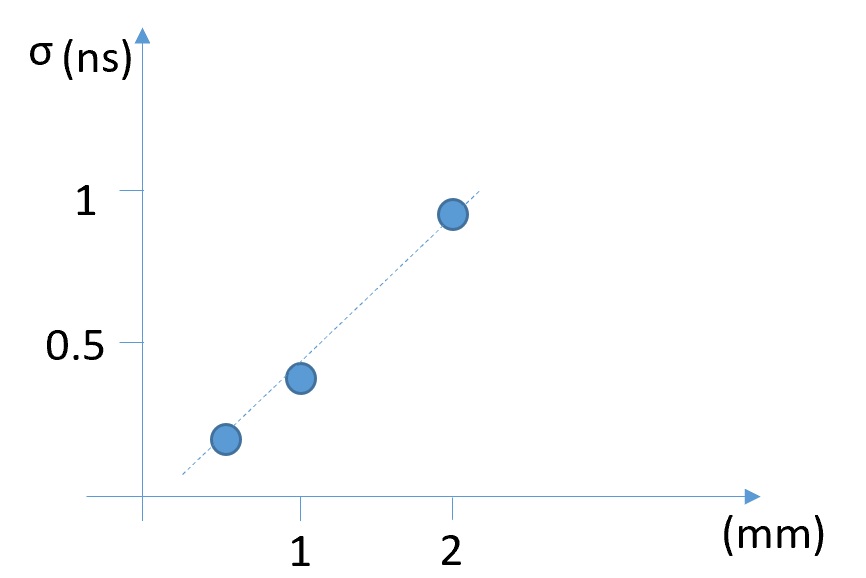}} \quad
\subfloat[][\emph{The prompt signal duration as function of the electric field for three different gas gap thicnesses.}]
{\includegraphics[trim=0.cm 0.cm 0cm 0.cm, clip=true,width=0.5\textwidth]{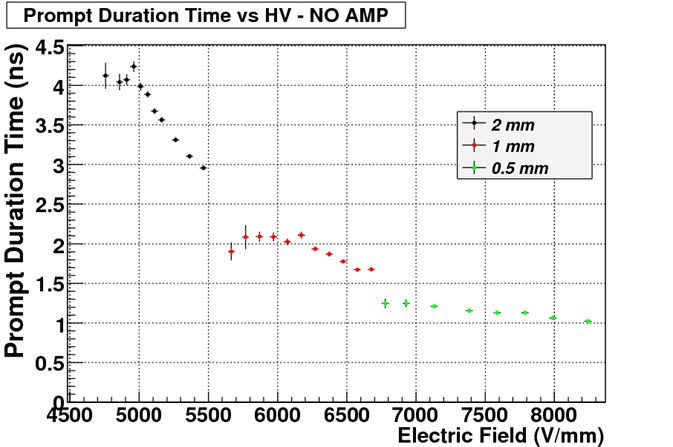}} \\
\caption{}
\label{fig:figure5}
\end{figure}

The RPC detector time resolution is an open question yet. Despite it is one of the most important features of this detector, the relations between the time jitter and the pulse shaping are unknown. What is well known is that the time resolution changes linearly with the gas gap thickness as figure \ref{fig:figure5}\emph{a} shows \cite{f}.
The prompt pulse duration time, showed in figure \ref{fig:figure5}\emph{b}, shows a saturation trend as the high voltage increase. This means that the electorns velocity fluctuations isn't the main factor that affect the time jitter. For this reason, at now the only way to improve the single gap time resolution is to thinning the gas gap. 

\section{Conclusions}
In this paper the connection between the front end electornics improvements and the detector developments is stressed. Respect to the first RPC, the new detectors boast a rate capability increase from $100\; Hz/cm^2$ to $10\;kHz/cm^2$. The time resolution is changed from $1\;ns$ to $30\;ps$ of the multigap detectors and the spatial resolution reached $100\;\mu m$. Nevertheless many aspects are still unknown, so the RPC can be considered a technology that has still a lot to give.

%



\begin{thebibliography}{99}

\bibitem{a}
{R. Santonico, R. Cardarelli}, \emph{NIM} {\bf 187} (1981) 377-380.

\bibitem{b}
{R. Cardarelli, R. Santonico}, \emph{NIM} {\bf 200} (1988) 263.

\bibitem{c}
\emph{ATL-COM-MUON-2017-033}, (July 7, 2017).

\bibitem{d}
{R. Cardarelli}, \emph{Scientifica Acta} {\bf 8} (1993) 159-165.

\bibitem{e}
{P. Camarri, R. Cardarelli, A. Di Ciaccio, R. Santonico}, \emph{NIM} {\bf 414} (1998) 317-324.

\bibitem{f}
{G. Aielli,... B. Liberti,... et al.}, \emph{Improving the RPC rate capability
}, \emph{Jinst} {\bf 11} (July 2016).

\bibitem{g}
{R. Cardarelli et al.}, \emph{Jinst} {10.1088/1748-0221/11/03/P03011} {IOP for SISSA Medialab} {9/2016}.





\end{thebibliography}
\end{document}